# Modeling the Dichotomy of the Immune Response to Cancer: Cytotoxic Effects and Tumor-Promoting Inflammation

Kathleen P. Wilkie and Philip Hahnfeldt

Center of Cancer Systems Biology, GRI,
Tufts University School of Medicine, Boston, MA, USA

**Running Title:** The Dichotomy of the Immune Response to Cancer

**Financial Support:** The work of K.W. and P.H. was supported by the National Cancer Institute under Award Number U54CA149233 (to L. Hlatky) and by the Office of Science (BER), U.S. Department of Energy, under Award Number DE-SC0001434 (to P.H.).

## Abstract

Although the immune response is often regarded as acting to suppress tumor growth, it is now clear that it can be both stimulatory and inhibitory. The interplay between these competing influences has complex implications for tumor development and cancer dormancy. To study this biological phenomenon theoretically we construct a minimally parameterized framework that incorporates all aspects of the immune response. We combine the effects of all immune cell types, general principles of self-limited logistic growth, and the physical process of inflammation into one quantitative setting. Simulations suggest that while there are pro-tumor or antitumor immunogenic responses characterized by larger or smaller final tumor volumes, respectively, each response involves an initial period where tumor growth is stimulated beyond that of growth without an immune response. The mathematical description is non-identifiable which allows us to capture inherent biological variability in tumor growth that can significantly alter tumor-immune dynamics and thus treatment success rates. The ability of this model to predict immunomodulation of tumor growth may offer a template for the design of novel treatment approaches that exploit immune response to improve tumor suppression, including the potential attainment of an immune-induced dormant state.

## Introduction

The role of the immune response in tumorigenesis is now generally accepted to be both stimulatory and inhibitory (Mantovani et al. 2008; de Visser et al. 2006). While the cytotoxic role of the immune system in tumor eradication has been known for centuries, it is only recently that the concept of immune stimulation of tumor development has become well accepted (de Visser et al. 2006; Hanahan & Weinberg 2011; Grivennikov et al. 2010; Rakoff-Nahoum 2006; Prehn 1972). One important mechanism of the pro-tumor immune response, inflammation, has been linked to tumor initiation (Kraus & Arber 2009), tumor progression (Mantovani et al. 2008; Grivennikov et al. 2010; Balkwill & Coussens 2004), and metastasis (Condeelis & Pollard 2006). Even inflammation, however, can be either stimulatory or inhibitory to tumor growth (Nelson & Ganss 2006).

The polarity of the tumor microenvironment, whether it be tumor-promoting or tumor-inhibiting, is determined by the intercellular interactions and cytokine signaling milieu. Tumor microenvironments include, amongst the extracellular matrix and stromal cells, a variety of innate immune cells (including macrophages, neutrophils, mast cells, myeloid derived suppressor cells, natural killer cells, and dendritic cells) and adaptive immune cells (including B and T lymphocytes). Each immune cell type may have both tumor-promoting and tumor-inhibiting actions. Macrophages, for example, can recognize and engulf cancer cells, but they can also promote tumor growth through the expression of cytokines and chemokines, which in turn stimulates angiogenesis (De Palma et al. 2005; Albini 2005), lymphangiogenesis (Ji 2011), and matrix remodeling (Condeelis & Pollard 2006). More generally, mechanisms of immune stimulation of cancer development include the induction of DNA damage by the generation of free radicals, the promotion of angiogenesis and tissue remodeling through growth factor, cytokine, chemokine, and matrix metalloproteinase production, the suppression of antitumor immune activities, and the promotion of chronic inflammation in the tumor microenvironment (de Visser et al. 2006). Mechanisms of immune inhibition of cancer development include the inhibition of tumor growth through direct cancer cell lysis, cancer cell apoptosis induced by perforin and granzymes or Fas/Fas-ligand binding, and the pro-inflammatory but antitumor production of cytokines such as IL-2, IL-12, and IFN-γ (Nelson & Ganss 2006). In fact, the shift of the inflammatory environment from pro-tumor and pro-angiogenic (with factors such as IL-4, IL-6, IL-10, and TGF-β) to antitumor and anti-angiogenic (with factors such as IL-2, IL-12, IP-10, MIG, and IFN-γ) may be crucial for tumor elimination, as even vascular endothelial cells have been shown to lyse target tumor cells once activated with TNF-α and IFN-γ (Li et al. 1991). Thus, the cytokines and growth factors present in the tumor microenvironment are not only crucial for the determination of the differentiated state of the immune response, but also for the determination of the response of local stromal cells to tumor presence.

Tumor microenvironments can be modified by the adaptive immune response and this effect can be enhanced by immunotherapy (Nelson & Ganss 2006). Cancer immunotherapy aims to improve tumor suppression by increasing cytotoxic strength. One type of cytokine immunotherapy involves the repeated injection of IL-2, which primes the T cell response with CD4+ and CD8+ T cells. With intratumoral injection of IL-2, tumor regression was shown to correlate with a reduction in tumor blood vessels and this process was shown to be dependent on functional CD8+ lymphocytes (Jackaman et al. 2003). The efficacy of this treatment, however, declined with increasing tumor size, which may be associated with a less effective immune response. The ability of the immune response to modify tumor vasculature suggests a potential anticancer strategy of cytokine-based immunotherapy, with inflammatory factors such as IFN-γ, to shift the pro-angiogenic and immunosuppressive tumor microenvironment to an anti-angiogenic microenvironment that supports cytotoxic immune activities. These cytotoxic activities could then be enhanced by cell-based immunotherapies such as adoptive T cell transfer or dendritic cell activation (Nelson & Ganss 2006). Figure 1 summarizes the significant immune cells and cytokines or growth factors involved in the pro-tumor and antitumor inflammatory responses.

To identify and track the complex mechanistic interplays underlying tumor development in the context of the immune response, we sought to distill the fundamental reciprocal interactions controlling the process into one quantitative framework. Heretofore, several approaches have been applied to quantify cancer-immune interactions. Perhaps the most common is to describe the system as a set of ordinary differential equations that capture the time-varying dynamics at the population level (DeLisi & Rescigno 1977; Kuznetsov et al. 1994; Kirschner & Panetta 1998; de Pillis et al. 2005; d'Onofrio 2005; d'Onofrio & Ciancio 2011; Eftimie et al. 2010; Wilkie 2013). Other approaches focus on random effects with stochastic differential equations (Lefever & Horsthemke 1978), spatio-temporal dependence with partial differential equations (Matzavinos et al. 2004; Roose et al. 2007; Al-Tameemi et al. 2012), and individual cell-cell interactions using agent-based methods (Takayanagi et al. 2006; Roose et al. 2007; Enderling et al. 2012). Among the several excellent reviews on the subject are those covering discrete tumor-immune competition approaches (Adam & Bellomo 1997), non-spatial, time-varying models (Eftimie et al. 2010), and analyses of the dormant or near-dormant tumor state (Wilkie 2013).

Despite the overwhelming evidence of direct immune stimulation of tumor growth, mathematical treatments of tumor-immune interactions have, until now, focused solely on the cytotoxic actions of immune cells. Some models do predict immune stimulation of tumor growth as a byproduct of cytotoxic inhibitory actions (Kuznetsov 1988; Joshi et al. 2009), but as we will show, explicit inclusion of stimulatory mechanisms, such as tumor-promoting inflammation, may be necessary to explain observations of tumor promotion prior to tumor suppression over the course of immunotherapy (Wolchok et al. 2009). We therefore propose a new mathematical framework that incorporates both immune-mediated tumor stimulation and tumor inhibition. This new model is capable of analyzing a more complete view of the interactions that occur between cancer cells and immune cells, and is therefore more suited for analysis and prediction of cancer treatment strategies, especially immunotherapies.

**Model Equations and Assumptions**
In this work, cancer and immune cell populations are assumed to grow according to a generalized logistic law that is mechanistically modified by their cellular interactions.

*Cancer Cell Population Growth*
The cancer population, $C(t)$, grows intrinsically up to a limiting size, the carrying capacity $K_C(t)$. Growth is inhibited by the immune system through predation, $\Psi$, which modulates the growth rate, and it is stimulated by the immune system through an inflammatory process incorporated in the carrying capacity. The cancer population is thus governed by

$$\frac{dC}{dt} = \frac{\mu}{\alpha}(1 + \Psi(I,C))C\left(1 - \left(\frac{C}{K_C}\right)^{\alpha}\right), \qquad C(0) = C_0. \tag{1}$$

The carrying capacity is determined by balancing stimulatory and inhibitory signals. Based on arguments of the relative clearance rates for these signals, it was suggested that the stimulation and inhibition terms for $dK_C(t)/dt$ be proportional to volume to the power 1 (in this case $C(t)^1$) and volume to the power of $\frac{5}{3}$ (in this case $K_C(t)^1 C(t)^{\frac{2}{3}}$), respectively (Hahnfeldt et al. 1999). Note that here we measure all compartments by cell number which is related to volume according to $1\,mm^3 = 10^6$ cells. Along these lines, we assume that pro-angiogenic signals are produced by both cancer and immune cells, but that their combined effects, which require cell-cell interactions, must be proportional to volume to the first power, i.e., $(B + I(t))^a C(t)^{1-a}$, where $B$ is a background constant enabling the cancer to stimulate its own growth in the absence of an immune response and $0 \le a \le 1$. In this formulation, pro-angiogenic signals resulting from cancer-immune interactions are appended to signals resulting from cancer-stroma interactions.

In a similar manner, cancer cells, immune cells, and the current microenvironment, $K_C(t)$, contribute to the anti-angiogenic signals in proportion to volume to the $5/3^{rd}$ power, with the cancer-immune interactions appended to the cancer-stroma interactions. The resulting expression is $K_C(t)^1 (B + I(t))^b C(t)^{\frac{2}{3}-b}$ with $0 \le b \le \frac{2}{3}$. Together, this mathematical formulation allows for pro- and anti-angiogenic signals to be produced by both cancer and immune cells. Defining $B = 1$ allows for cancer growth without any extrinsic stimulation, and thus the cancer carrying capacity is governed by

$$\frac{dK_C}{dt} = p(1+I)^a C^{1-a} - qK_C(1+I)^b C^{\frac{2}{3}-b}, \qquad K_C(0) = K_{C,0}. \tag{2}$$

We assume that the majority of these signals should originate from the cancer population and thus require $a$ and $b$ to be small. Parameter $a$ controls the weight of the immune population's contribution to tumor-promoting factors (such as pro-angiogenic growth factors) that act to increase the tumor's carrying capacity. Parameter $b$ controls the weight of immune contributions to tumor-inhibiting factors, (such as anti-angiogenic growth factors) that act to limit the tumor's carrying capacity. Cytotoxic T cell activity, Th1 immunity, and cytokines such as IFN-γ, IL-2, and IL-12 are significant factors that contribute to an anti-angiogenic tumor-inhibiting inflammatory microenvironment (Nelson & Ganss 2006). A pro-angiogenic tumor-promoting inflammatory microenvironment is associated with immunosuppressive myeloid cells, Th2 immunity, and cytokines such as TGF-β, IL-4, IL-6, and IL-10 (DeNardo et al. 2010). See Figure 1.

When $a > b$, more weight is placed on the tumor-promoting effects of immune cells and we label this case as *pro-tumor immunity*. When $a < b$, more weight is placed on the tumor-inhibiting effects of immune cells and we label this case as *antitumor immunity*. In this work, we chose $a = \frac{2}{10}$, which allows immune cells to contribute to tumor promotion but requires the majority of stimulation to originate from the tumor itself. We then choose $b = \frac{1}{10}$ (a value slightly less than $a$) for pro-tumor immunity and $b = \frac{3}{10}$ (a value slightly more than $a$) for antitumor immunity.

*Immune Cell Population Growth*

The immune population, $I(t)$, which includes all various immune cell types, maintains a homeostatic (equilibrium) state, $I_e$, unless stimulated to grow in response to the cancer's presence through direct recruitment, $rC$, and through cancer-immune interactions that increase the carrying capacity, $K_I(t)$. Thus, the immune population is governed by

$$\frac{dI}{dt} = \lambda \left( I + rC \right) \left( 1 - \frac{I}{K_I} \right), \qquad I(0) = I_0. \tag{3}$$

Note that we simplify generalized logistic growth to logistic growth here to reduce the number of immune system parameters. The immune population carrying capacity is governed by

$$\frac{dK_I}{dt} = xI^{\frac{1}{2}}C^{\frac{1}{2}} - yK_I I^{\frac{1}{3}}C^{\frac{1}{3}} - z\left( K_I - I_e \right), \qquad K_I(0) = K_{I,0}. \tag{4}$$

On the right hand side, the first two terms prescribe equal weight to the two populations in determining how the immune carrying capacity grows in response to stimulatory and inhibitory signals. Here we chose the powers of each term to match the dynamics of the cancer carrying capacity. Note that immune system signals must act both locally and systemically so the clearance rates of both stimulatory and inhibitory cytokines are necessarily variable. Control of the immune response is obtained, however, through the actions of checkpoint blockades and regulatory T cells, as well as other mechanisms, and thus is determined by cancer-immune interactions. The last term represents an organismic tendency of the immune response to return to a healthy homeostatic state after disease elimination.

*Immune Predation of Cancer Cells*

Direct cytotoxic immune actions targeted against cancer cells are modeled by

$$\Psi = -\theta \left( \frac{I^{\beta}}{\phi C^{\beta} + I^{\beta}} + \epsilon \log_{10}(1 + I) \right), \tag{5}$$

where the term $\dfrac{I^{\beta}}{\phi C^{\beta} + I^{\beta}}$ describes the saturation kinetics of strong cytotoxic actions (de Pillis et al. 2005) and the term $\epsilon \log_{10}(1 + I)$ allows for a gradual increase to this saturation level with significant increase in immune cell number. The ratio-dependent saturation term was shown to describe the cytotoxic effects of T cells (de Pillis et al. 2005), but this neglects the contribution from innate immunity (natural killer cells and macrophages) which do not exhibit saturation in cytotoxic assays (Diefenbach et al. 2001). The logarithmic term accounts for innate immunity at large population sizes and phenomenologically maps the actions into a range appropriate for $\Psi$. For small immune populations, innate and adaptive predation can be combined into the ratio-dependent term, but for large populations, innate immunity should still have an effect (here we assume a small effect and set $\epsilon = 0.01$). Without an increasing predation threshold, tumor growth dynamics, especially after periods of immune-induced tumor dormancy, would not reflect the still growing immune presence, which could have significant implications on immunotherapy predictions. Thus, this form accounts for adaptive and innate cytotoxic effects over a wide range of immune population sizes, which are both required for tumor elimination (Koebel et al. 2007).

**Experimental Data**

Equations (1)-(4) govern a system of four dependent variables, $C(t)$, $I(t)$, $K_C(t)$, and $K_I(t)$, which describe the growth dynamics of a tumor in the presence of a complete and competent immune system capable of both stimulating and inhibiting tumor growth. Parameters describing tumor growth and immune predation are estimated from the following datasets.

Basic tumor growth kinetics are estimated from experimental measurements of a subcutaneous fibrosarcoma induced by 3-methylcholanthrene in wild-type mice (Tanooka et al. 1982). Measurements were taken once the resulting tumors reached a palpable size.

Stimulatory and inhibitory effects of specifically-trained immune cells (cells that were previously exposed to the specific cancer) on tumor growth has been investigated (Prehn 1972). Varying numbers of splenocytes were mixed with fibrosarcoma cells (from tumors induced by 3-methylcholanthrene) and injected into mice that were thymectomized as adults and whole-body irradiated 24 hours before injection to neutralize host immunity). Specifically-trained spleen cells were collected from syngeneic mice that had grown the same tumor for 10-20 days, then left to recover for 7-12 days after excision of the tumor. Varying numbers of splenocytes were mixed with $10^4$ sarcoma cells, injected subcutaneously, and allowed to grow until the largest tumor reached a diameter of 10 mm. Data suggests that specifically-trained immune cells stimulate tumor growth when mixed at ratios smaller than parity and inhibit tumor growth when mixed at ratios larger than parity. This dose-response curve was idealized to a parabolic shape with a maximum located at cancer-immune parity (Prehn 2007).

**Parameter Estimation**

Tumor growth parameters are estimated in stages using a Markov Chain Monte Carlo (MCMC) method (Robert & Casella 2010; Cirit & Haugh 2012). From an initial guess, a Markov chain of permitted parameter sets is created by randomly perturbing the previous parameter set and accepting this perturbed set with a probability determined by a measure of the goodness of fit, here the sum of squared deviations. Each parameter is perturbed and tested for acceptance independently, except parameters $\mu$ and $\alpha$, which are perturbed and tested together since they are inherently related in equation (1).

Parameters $\mu$, $\alpha$, $p$, $q$, $K_{C,0}$, and $I_e$ are estimated from fibrosarcoma tumor growth data (Tanooka et al. 1982). Such tumors grown in wild-type (immune competent) mice are nonimmunogenic due to early immunoediting (Cohen et al. 2010) and regulatory T cells can inhibit immune-mediated tumor rejection (Betts et al. 2007). We therefore assume that growth after the tumor reached the palpable size of about $1.3 \cdot 10^7$ cells occurred primarily in a host where both immune recruitment and immune predation was negligible. This simplifying assumption allows us to set $\frac{\mathrm{d}I}{\mathrm{d}t} = \frac{\mathrm{d}K_I}{\mathrm{d}t} = 0$ in equations (3) and (4), maintaining immune presence at the homeostatic level $I(t) = K_I(t) = I_e$, and to set $\Psi = 0$ in equation (1), enforcing zero immune predation. This assumption underestimates the role of immune stimulation in tumor growth, but we accept this limitation in favor of simplifying the parameterization procedure. To estimate the tumor growth parameters, our MCMC method was run 10 times with 20,000 trials per parameter in each run. Fitting equations (1) and (2) with $a = \frac{2}{10}$, $b = \frac{3}{10}$, $C_0 = 1.3 \cdot 10^7$, and $I(t) = K_I(t) = I_e$, to the experimental growth data gives the 10 parameter sets for antitumor immunity listed at the top of Table 1. Fitting the same equations to the data with $b = \frac{1}{10}$ gives the 10 parameter sets for pro-tumor immunity listed at the bottom of Table 1.

This approach results in two ensembles of 10 parameter sets, one for antitumor immunity and one for pro-tumor immunity. Due to limited experimental data and the necessary level of complexity in our mathematical model, some parameter values are non-identifiable, as seen through the variability of the 10 sets. The problem of parameter identifiability is becoming increasingly important, especially in the growing areas of nonlinear ODE modeling with applications to biological networks and immune models for viral infections and cancer. It may be resolved by measuring or acquiring additional data (which may not be possible) or by model reduction, techniques of which are still under development (Miao et al. 2011; Meshkat et al. 2011; Raue et al. 2011). Our approach is to instead generate an ensemble of parameter sets that represent 10 individual patients with their own inherent variabilities instead of an "average responder", and to use this ensemble to better explore the implications of cancer-immune interactions on tumor growth dynamics for a population of individuals.

Our assumption of no immune recruitment to the tumor site is an oversimplification. Since these tumors grow in wild-type mice from carcinogen injection, however, based on observations (Cohen et al. 2010; Betts et al. 2007) we can assume that the immunoediting phase of tumor progression occurs prior to visible detection. The first tumor measurement is thus taken as the initial volume for the parameter-fitting algorithm, and the tumor is assumed to be negligibly immunogenic, stimulating no significant immune response.

To estimate immune predation parameters we use experimental data for tumor growth resulting from co-injections of specifically-trained immune cells and fibrosarcoma cells (Prehn 1972). In these experiments, mice were subjected to whole body irradiation and thymectomized so that host immunity is negligible. Thus, the injected immune cells are the only immune cells present in the system; their actions may be stimulatory or inhibitory to tumor growth, but we assume that no immune recruitment or proliferation may occur. Again, this allows us to simplify the equations by setting $\frac{\mathrm{d}I}{\mathrm{d}t} = \frac{\mathrm{d}K_I}{\mathrm{d}t} = 0$ in equations (3) and (4). To estimate the parameters for direct cytotoxic effects of the immune response, $\theta$, $\phi$, and $\beta$ from equation (5) were manually tuned (via trial and error over a broad range of values) until the maximal tumor stimulation occurred near cancer-immune parity and tumor elimination occurred with sufficient immune presence. The resulting parameter estimates are listed in Table 2.

The final stage of parameterization involves estimating parameters that describe immune growth and recruitment in the wound healing process. Due to a lack of experimental data, we estimate these parameters based on the assumption that the immune response should grow approximately as fast and as large as the tumor mass. These parameter values are listed in Table 2.

**Numerical Simulations and Results**

Ten parameter sets are estimated using an MCMC method for both the antitumor and pro-tumor immunity cases, see Table 1. Significant variability exists in the numerical values for each parameter and yet each set fits the experimental data equally well (compare the sum of squared deviations listed for each set in Table 1). This parameter variability can cause significant changes in the phase-space (cancer-immune dynamics) of the model, as shown by the phase portraits in Appendix A1 (antitumor) and A2 (pro-tumor). Figure 2a shows the phase-space dynamics for an average and an outlier parameter set from both the antitumor and pro-tumor immunity cases.

The functional form of equation (2), which describes the growth of the cancer population carrying capacity, incorporates effects of both immune-mediated stimulation and inhibition of tumor growth. Figure 2b shows the effect of constant immune presence on tumor growth in both the antitumor ($a < b$) and pro-tumor ($a > b$) inflammatory environments without direct predation ($\Psi = 0$). Under this assumption, antitumor immunity stimulates tumor growth but also limits the final tumor burden. That is, the tumor may progress and grow faster with immune stimulation, but the maximum obtainable size is ultimately reduced. In fact, the amount of early-stimulation and late-inhibition of tumor growth arising from immune activities is sensitive to the tumor growth parameters, and thus is a feature inherent to the individual tumor, the tumor's microenvironment, and the host. When protumor immunity is assumed, however, more weight is placed on the pro-angiogenic activities of inflammation and, as a result, the tumors are predicted to grow faster and larger than those predicted to grow without immune presence.

The pro-tumor and antitumor cases present two fundamentally different classes of possible outcomes: immunomodulation causing tumors to grow faster but be ultimately smaller, or faster and ultimately larger, than those growing in the absence of an immune response. When $a = b$ in equation (2), immune-stimulation of tumor growth is predicted, causing the tumors to grow faster in the presence of immune cells, but the ultimate tumor size is fixed and independent of immune presence.

Dose response curves for both pro-tumor and antitumor immunity are shown in Figure 2c. Tumors are simulated to grow with varying initial numbers of immune cells, predation is included but no immune growth occurs. Both an average and an outlier parameter set are shown for each type of immunity. The dose response at three different time points are compared to the experimental data (Prehn 1972). Average and outlier parameter sets behave differently (hence the different time points used), but all predict immune-mediated stimulation and inhibition of tumor growth according to a parabolic shape. Importantly, the model is able to predict the ratio-dependence of tumor stimulation by immune cells, as observed by Prehn. That is, immune stimulation of tumor growth occurs when cancer cells outnumber immune cells and inhibition occurs when immune cells outnumber cancer cells.

As the phase portraits in Figure 2a demonstrate, ultimate tumor fate is determined by the initial conditions of this deterministic model. Biologically, this may translate to the level and polarization of the immune response once the cancer reaches a critical size. Biological determinants of immune presence when the critical cancer size is reached may include the antigenicity of the cancer cells, which is related to their accumulated mutations, and the location of the cancer, as different tissues have different levels of immune surveillance. To compare our simulations, we say that this critical size is the same as our initial cancer size, $C_0 = 10^4$ cells, and thus relate the initial immune presence to this value via $I_0 = \gamma C_0$, where $\gamma$ is a constant. Each parameter set has a different threshold value for $\gamma$, wherein immune presence less than $\gamma C_0$ results in tumor escape and that greater than $\gamma C_0$ in elimination. Before these dichotomous outcomes are achieved, however, seemingly contradictory events may occur, such as a transient period of dormancy prior to tumor escape, or growth and stimulation prior to elimination.

Pro-tumor inflammatory environments likely have immunosuppressive mechanisms that reduce predation efficacy. Such mechanisms, not yet considered by this model, may alter the predation parameters $\phi$, $\beta$, and $\theta$, and thus the dose response curves predicted here. Furthermore, this framework does not allow for the evolution of antitumor immunity into pro-tumor immunity during tumor development. We note that excluding these immunosuppressive and immunoevasive mechanisms are limitations of our model and leave them to future work.

For improved tumor control, the polarization of the microenvironment should be shifted from pro-tumor to antitumor immunity via immunotherapies. We thus focus on antitumor immunity for parameter sensitivity using a parameter set demonstrating typical behavior (set 5 from Table 1). Immunotherapies that enhance the antigenicity of cancer cells can be incorporated into the model through the immune recruitment parameter $r$. Increases in recruitment result in improved tumor suppression and a reduction in tumor stimulation, as shown in Figure 3.

Effects of variations in homeostatic regulation, $z$, are also shown in Figure 3. As resistance to altering the immune homeostatic state is increased, tumor suppression is reduced and immune-mediated dormancy periods may become more difficult to achieve. As seen in Figure 3c, increased immune resistance slows immune growth, requiring higher initial immune presence for elimination. Thus, with increased homeostatic strength, dormant tumors are smaller and require more initial immune presence to be obtained. Biologically, this may translate into periods of tumor dormancy for only highly immunogenic cancers in hosts with large homeostatic resistances, a parameter that may change with host health and age. Homeostatic regulation is an intrinsic patient-specific parameter that is often neglected in mathematical models, however, as is demonstrated here, it may significantly affect tumor growth dynamics, and in particular, tumor dormancy. Increased immune recruitment, on the other hand, seems to improve tumor suppression while maintaining the ability to achieve a dormant tumor state, Figure 3d.

Contour plots for parameter sensitivity are shown in Figure 4 for the following parameter pairs: homeostasis and recruitment $(r,z)$, immune growth rate and predation efficacy $(\lambda,\theta)$, and predation shape parameters $(\beta,\phi)$. For each pair four time points are shown. After 150 days, most of the predicted tumor fates have already been established. That is, either the tumor has been eliminated or escaped. Of interest, are the intermediate regions as they represent tumors whose fates may be altered through immunotherapies to achieve elimination or prolong the dormant state. The fact that immune-induced dormancy is observed in several experimental models (Quesnel 2008) suggests that parameters predicted to modify this state, such as homeostatic strength or immune recruitment, may be desirable targets for immunotherapy. In Figure 4, the intermediate shades are intervention windows, or optimal ranges for parameter values to affect tumor fate. Since these windows shrink considerable over time, combination therapies may be necessary to re-open the window for immune intervention.

Figure 5 demonstrates parameter sensitivity for immune-related model parameters on atypical tumor growth behavior (antitumor parameter set 9 from Table 1). With this parameter set, the model does not predict significant dormant periods prior to elimination. For immune recruitment and homeostasis parameters, larger changes are required to see an alteration in tumor fate compared to set 5. An increased immune growth rate of $\lambda = 0.4$, for example, may delay tumor growth, but not alter the fate (compare contour plots at 50 and 100 days). Comparing Figures 3 and 4 to Figure 5 demonstrates that treatment outcomes depend on the intrinsic growth dynamics that are captured here by the parameter ensembles. For example, a treatment intended to increase immune recruitment based on successful predictions of tumor elimination from Figure 3 with parameter set 5 (say to the level of $r = 1$), may not alter tumor fate at all for parameter set 9 from Figure 5.

**Discussion**

To investigate the role of tumor-promoting inflammation, an emerging hallmark of cancer (Hanahan & Weinberg 2011), a new mathematical model for cancer-immune interactions was presented. This framework captures both pro-angiogenic, tumor-progressing actions of pro-tumor immunity, and anti-angiogenic, tumor-inhibiting actions of antitumor immunity. The use of generalized logistic growth captures some of the inherent variability underlying tumor growth dynamics in an immune competent host, often neglected in macroscopic measurements and mathematical models. Model simulations suggest that two types of inflammatory responses (pro-tumor or antitumor) resolve into two fundamentally different classes of outcomes, where inflammation-enhanced tumor progression results in either a decreased tumor burden, as in the anti-tumor case, or an increased tumor burden, as in the pro-tumor case. Thus, near- and long-term responses of a tumor to immune interaction may be opposed; that is, a response dynamic that appears to promote growth in the near term may be superior at curtailing growth in the long-term, even to the point of establishing dormancy, while the other allows for tumor escape. These results suggest that, in some cases, stimulated tumor growth early on may be advantageous, if it leads to a significantly smaller tumor burden. In such cases, treatments may be targeted to enhance the stability of an anti-tumor inflammatory environment instead of immediate tumor regression.

A Markov chain Monte Carlo method was used to estimate parameter sets that predict tumor growth equally well, but that, at the same time, also predict fundamentally different underlying dynamics. The results underscore the ultimately polar nature of final tumor fate (escape or elimination), and, at the same time, indicate that persistent regions of near-dormancy may precede either of these two outcomes. The striking variability observed across the parameter sets demonstrates the significance of intrinsic and immeasurable factors determining the complex biological processes involved in tumor growth in an immune competent host. The underlying variability in tumor dynamics, often neglected by mathematical formulations, is captured here by generalized logistic growth with a dynamic carrying capacity.

We propose that this variability, which is not measurable through macroscopic observations, may be due to the sensitivity of cancer cells and host stromal cells to growth and regulatory signals present in the microenvironment. Biological contributors to this variability may include the response rate of the host to pro- or anti-angiogenic signals or the strength of (or sensitivity to) the size-limiting signals originating from the tumor microenvironment (carrying capacity). Consequently, if treatment strategies are designed based on an average behavior parameter set (or patient), the treatment cannot be expected to result in the same outcome for all parameter sets (or patients). In fact, this variability may explain why treatments, including immunotherapies, work for some cases, but not all, and it emphasizes the importance of patient-specific treatment planning.

This quantitative framework also demonstrates an important and often oversimplified feature of tumor dormancy, that dormancy is a transient state. Many mathematical models predict dormancy as a stable equilibrium solution that is attained and maintained for infinite time (Wilkie 2013). The model presented here, however, describes dormancy as a transient phase that exists between tumor elimination and tumor escape, and it suggests that while treatments may prolong this state, by the fundamental nature of dormancy, the period must eventually transition to either elimination or escape.

Immunotherapies, which aim to boost patient immune responses to control or eliminate the disease, have met with some success, but have failed to produce a broadly effective treatment option (Phillips 2012). The immune-stimulating drug Levamisole, for example, has been reported to inhibit tumor growth at low doses but to have no inhibitory effect, compared to control, at high doses (Sampson et al. 1977). This dose-response may result from small doses enhancing the antitumor immune response while large doses over stimulate the response, promoting a conversion from antitumor to pro-tumor immunity, and ultimately enhancing tumor development. Such hypotheses highlight the need for theoretical investigation of dose-response and dose-scheduling in treatment planning, which can be performed with our proposed framework since both direct tumor-inhibiting and direct tumor-promoting mechanisms of the immune response are considered. The model simulations and results discussed here suggest that key factors for improved tumor control by immunotherapies include an understanding (and incorporation) of patient-specific inherent variability in tumor growth dynamics, consideration of the type of immune response active within the tumor microenvironment (pro-tumor versus antitumor), and optimal treatment dosages and schedules (the subject of ongoing work).

## Acknowledgements

The authors would like to thank Dr. M. La Croix for his helpful discussions and for creating Figure 1. The work of K.W. and P.H. was supported by the National Cancer Institute under Award Number U54CA149233 (to L. Hlatky) and by the Office of Science (BER), U.S. Department of Energy, under Award Number DE-SC0001434 (to P.H.). The content is solely the responsibility of the authors and does not necessarily represent the official views of the National Cancer Institute or the National Institutes of Health.

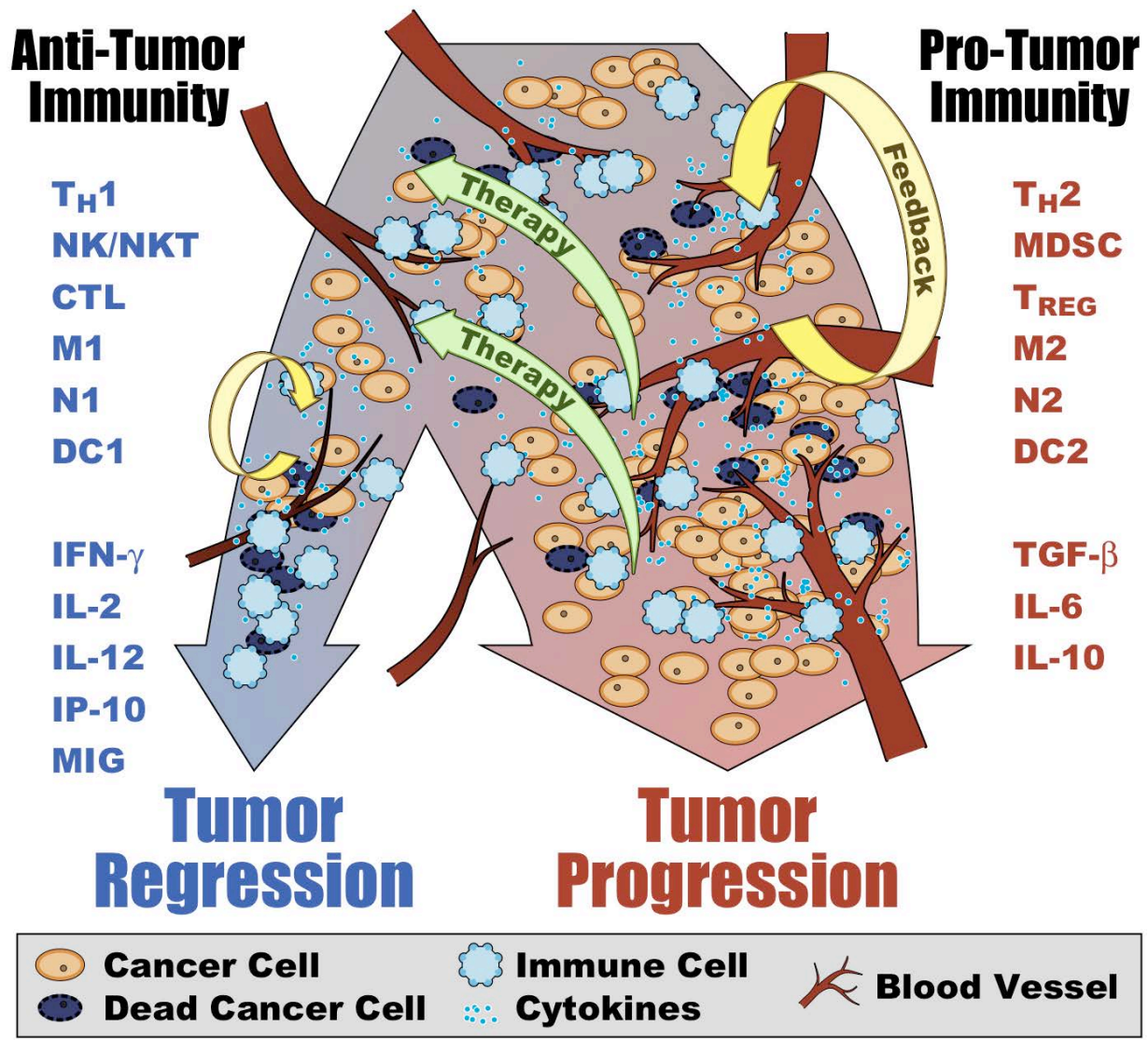

**Figure 1.** Pro-tumor and antitumor inflammatory responses are composed of different (or differently polarized) immune cells and cytokines / growth factors. The cytokine milieu present in the environment determines the polarization of newly differentiating immune cells. The cytokines produced by these new cells allows for a strong feedback mechanism to enhance the polarity of the immune response. Over time, these mechanisms may lead to the development of either a pro-tumor immunity that enhances vascularization and tumor progression, or an antitumor immunity that reduces vascularization and enhances tumor regression. Therapies that target the immune cells and cytokines present in the environment attempt to shift a pro-tumor immunity to an antitumor immunity to improve tumor suppression.

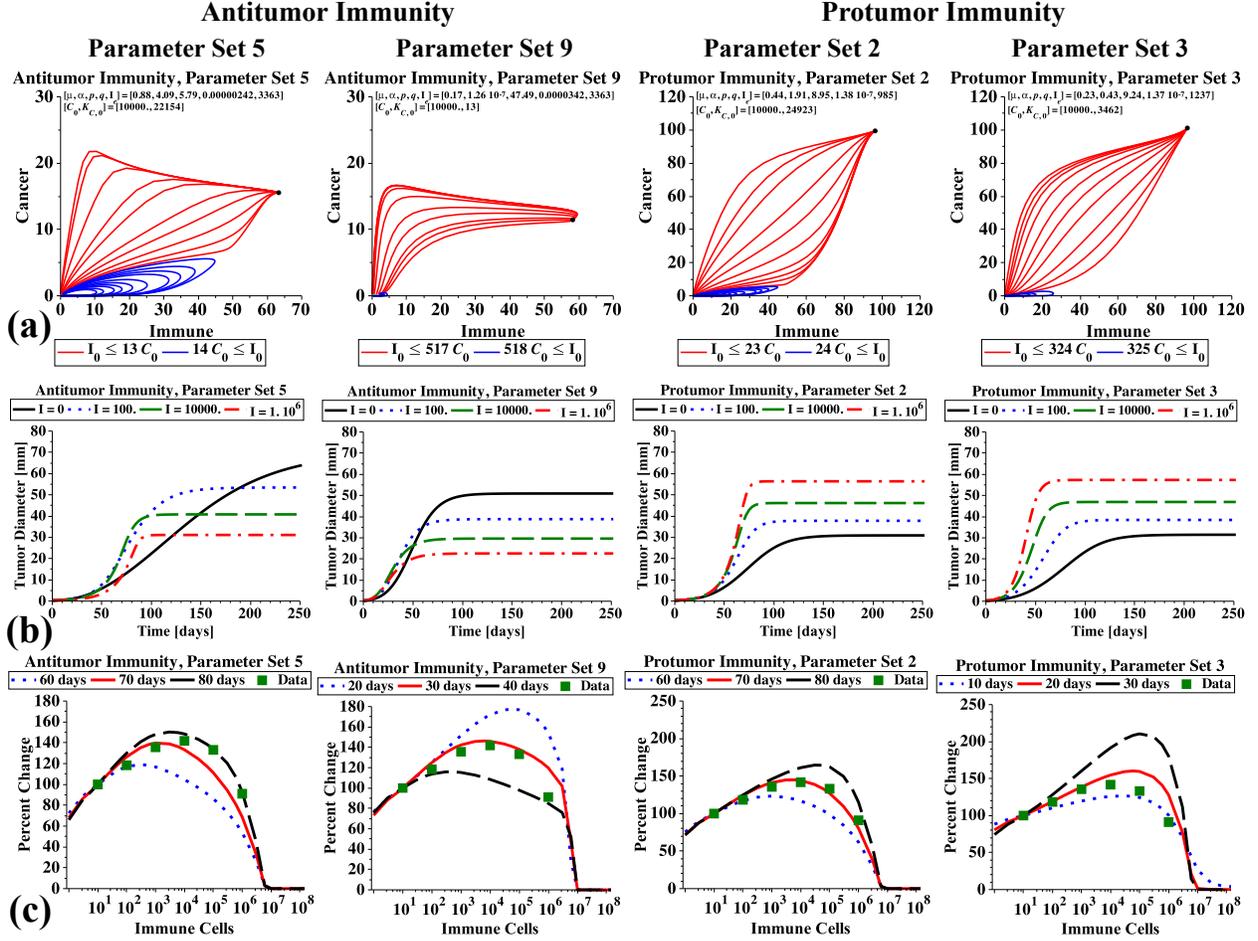

**Figure 2.** The behavior of antitumor and protumor immunity with both a typical and atypical parameter set. Phase portraits **(a)** demonstrate cancer-immune dynamics. The effect of parameter variability is demonstrated by the striking differences apparent between these phase portraits. Tumors are simulated to grow from an initial injection of $C_0 = 10^4$ cancer cells and varying numbers of immune cells ($I_0 = \gamma C_0$). The ranges (values of $\gamma$) that divide the behavior between tumor growth (red) or tumor suppression (blue), are listed below the plots. Simulations result from solving the full system of equations (1)-(4) with direct predation through $\Psi$, equation (5). Axes indicate diameter of spherical population in mm. Immune stimulation with **(c)** and without **(b)** predation is demonstrated. In **(b)** tumors are simulated to grow from an initial injection of $10^4$ cancer cells mixed with 0 (solid black), 100 (dotted blue), $10^4$ (dashed green), or $10^6$ (dash-dotted red) immune cells. Simulations result from solving equations (1) and (2) with $I(t)$ and $K_I(t)$ constant and $\Psi = 0$. In **(c)**, tumors are simulated to grow from an initial injection of $10^4$ cancer cells mixed with varying numbers of immune cells ranging from 0 to $10^8$. Three time snapshots of the dose-response (in terms of percent change) are shown in each plot along with the experimental data (Prehn 1972). Simulations result from solving equations (1) and (2) with $I(t)$ and $K_I(t)$ constant and $\Psi$ given by equation (5).

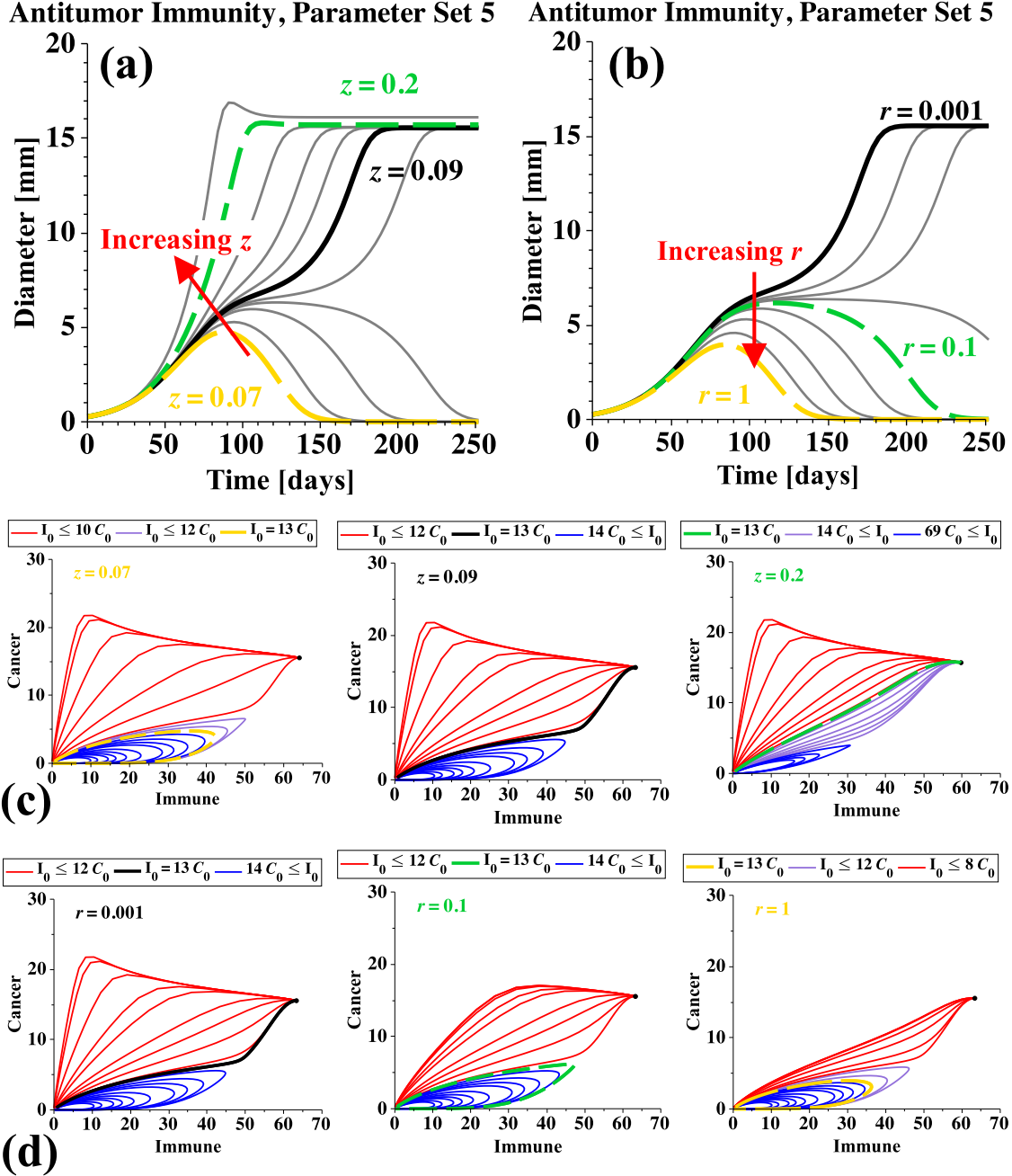

**Figure 3.** Parameter sensitivity for the homeostasis parameter $z$ (**a**) and the immune recruitment parameter $r$ (**b**) with the corresponding phase portraits (**c** and **d**) for antitumor immunity ($a < b$) and parameter set 5 from Table 1. Phase portraits are shown in (**c**) for three different values of the immune homeostasis parameter $z$ corresponding to the highlighted curves in (**a**). Increasing immune homeostatic resistance results in reduced tumor suppression and possibly a decreased ability to achieve tumor dormancy. Phase portraits are shown in (**d**) for three different values of the immune recruitment parameter $r$ corresponding to the highlighted curves in (**b**). Increasing immune recruitment results in a reduction of tumor stimulation and improved tumor suppression.

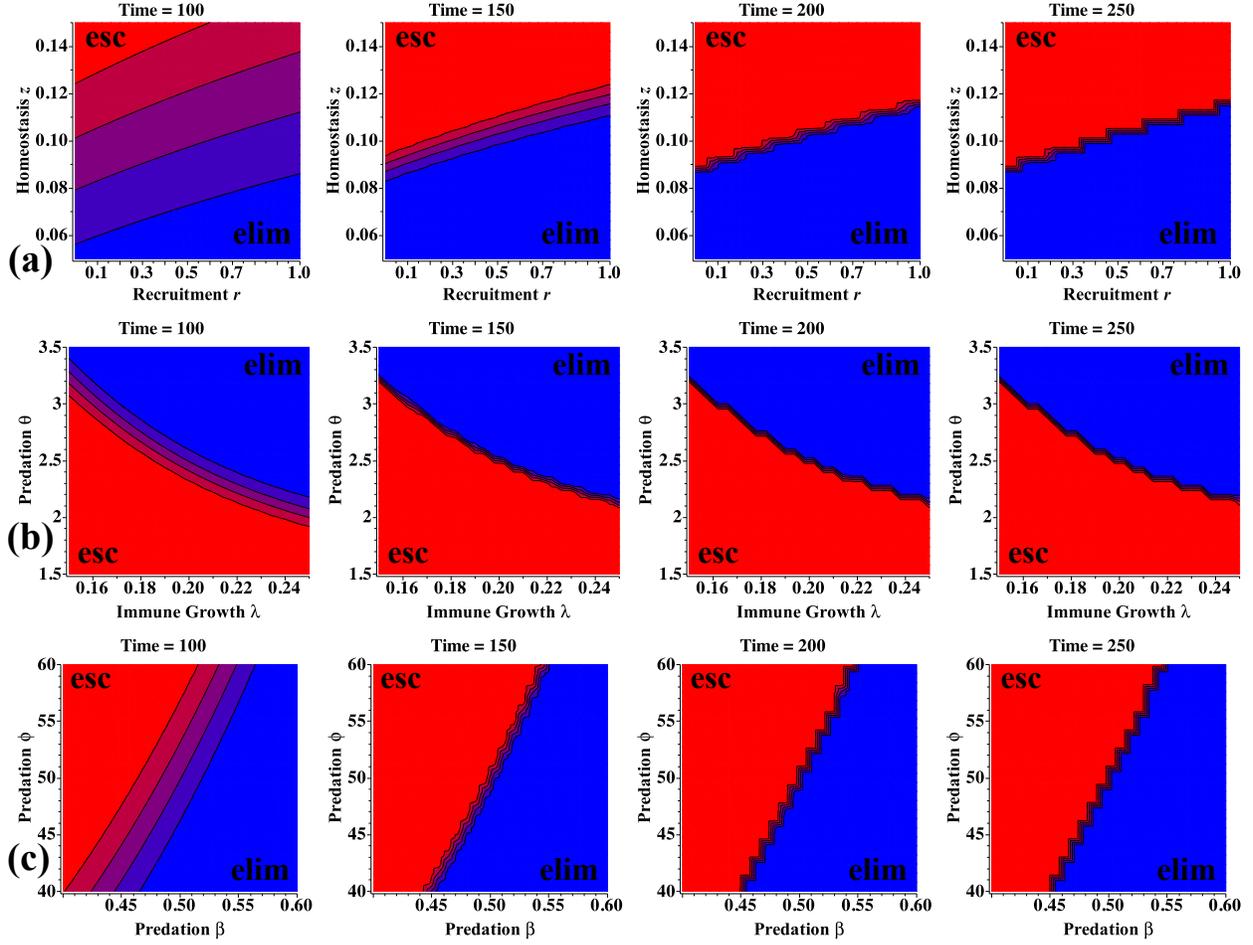

**Figure 4.** Parameter sensitivity contour plots of tumor fate for the parameter pairs $(r,z)$, $(\lambda,\theta)$, and $(\beta,\phi)$ are shown for antitumor immunity $(a<b)$ with parameter set 5 from Table 1. Contour plots at various times demonstrate the dependence of tumor fate on parameter values. Blue (elim) corresponds to tumor elimination and red (esc) to tumor escape. Shades of purple in between these limits correspond to tumors of intermediate size at the given time. These intermediate contour bands (especially at 100 and 150 days in **(a)** and at 100 days in **(b)** and **(c)**) indicate that tumor fate is susceptible to intervention early on, but that this window closes rapidly with time. This suggests that immunotherapies that modulate these parameters may need to be combined with alternate treatments to prolong these intervention windows.

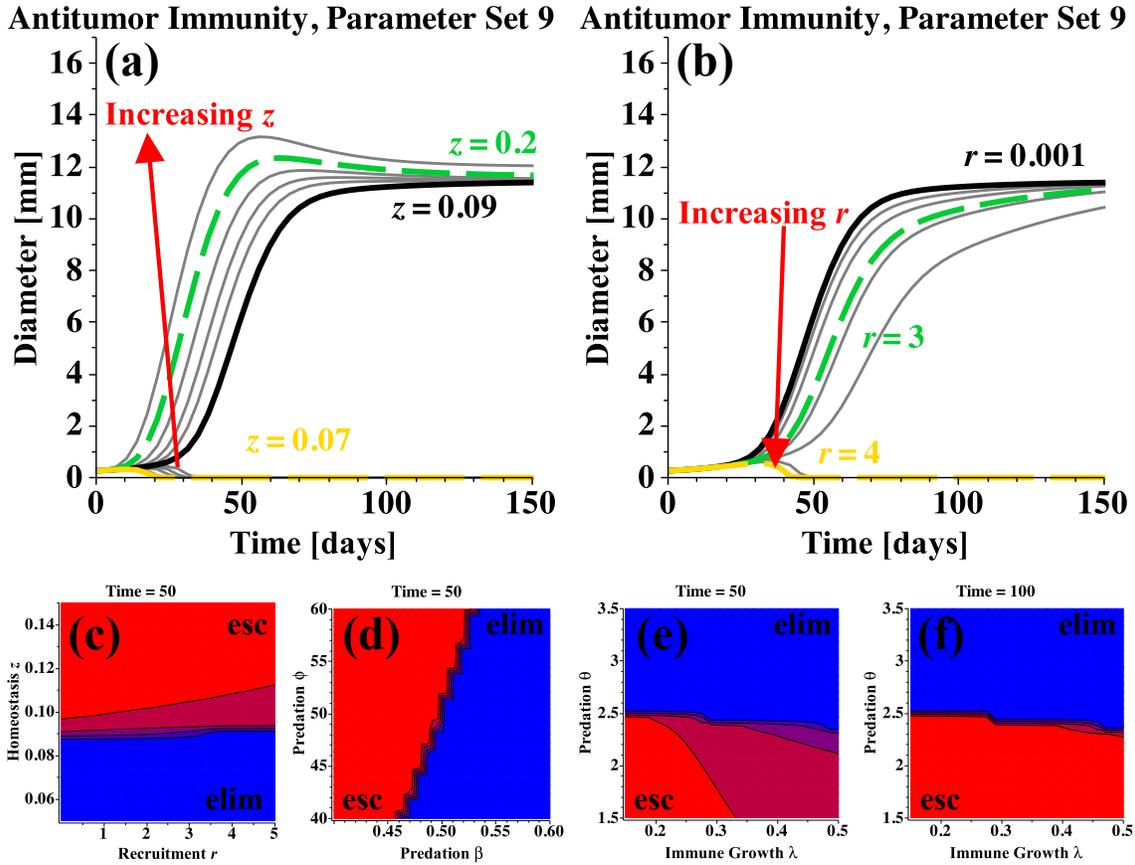

**Figure 5.** Parameter sensitivity for antitumor immunity ($a < b$) and the outlier parameter set 9 from Table 1. Tumor fate is altered as the homeostasis **(a)** and immune recruitment **(b)** parameters are increased. No significant dormant periods are predicted. Contour plots of tumor fate for the parameter pairs $(r,z)$, $(\beta,\phi)$, and $(\lambda,\theta)$ are shown in **(c)**, **(d)**, and **(e)** at 50 days, as well as $(\lambda,\theta)$ at 100 days in **(f)**.

**Table 1.** Antitumor and pro-tumor immunity parameter sets resulting from 10 runs of the MCMC parameter estimation method (at $20000$ iterations) which fit equations (1) and (2) with $b = \frac{3}{10}$ (antitumor) or $b = \frac{1}{10}$ (pro-tumor) and $I(t) = I_e$, a constant, to the tumor growth data of Tanooka *et al.*(Tanooka et al. 1982) using the initial guess of $(\mu, \alpha, p, q, K_{C,0}, I_e) = (1,1,1,10^{-6}, 3C_0, 10^3)$. The goodness of fit is measured by the sum of squared deviations (SSD) and the parameter set giving the minimum SSD is listed. Each parameter set fits the data equally well. Note that $K_{C,0}$ is scaled appropriately when $C_0$ is changed in the simulations to reflect different experimental conditions.

| | Parameter Set | Minimum SSD | $\mu$ [day$^{-1}$] | $\alpha$ [1] | $p$ [day$^{-1}$] | $q$ [[Cell No.]$^{-2/3}$ day$^{-1}$] | $K_{C,0}$ [Cell No.] | $I_e$ [Cell No.] |
|---|---|---|---|---|---|---|---|---|
| **Antitumor Immunity** | 1 | 3.66 | 0.29 | 1.31 | 111.38 | $7.85 \cdot 10^{-5}$ | $4.51 \cdot 10^6$ | 28.50 |
| | 2 | 3.89 | 0.34 | 1.46 | 20.07 | $9.54 \cdot 10^{-6}$ | $9.11 \cdot 10^6$ | 1294 |
| | 3 | 4.00 | 0.29 | 1.23 | 18.39 | $8.52 \cdot 10^{-6}$ | $4.54 \cdot 10^6$ | 7612 |
| | 4 | 7.20 | 8.81 | 42.63 | 4.61 | $1.86 \cdot 10^{-6}$ | $3.02 \cdot 10^7$ | 1850 |
| | 5 | 4.81 | 0.88 | 4.09 | 5.79 | $2.42 \cdot 10^{-6}$ | $2.88 \cdot 10^7$ | 3363 |
| | 6 | 3.53 | 0.43 | 1.94 | 10.81 | $4.72 \cdot 10^{-6}$ | $4.55 \cdot 10^7$ | 3635 |
| | 7 | 3.76 | 0.53 | 2.44 | 7.39 | $3.02 \cdot 10^{-6}$ | $4.09 \cdot 10^7$ | 4964 |
| | 8 | 3.75 | 0.59 | 2.73 | 7.53 | $3.19 \cdot 10^{-6}$ | $5.11 \cdot 10^7$ | 3993 |
| | 9 | 12.78 | 0.17 | $1.26 \cdot 10^{-7}$ | 47.49 | $3.42 \cdot 10^{-5}$ | $1.72 \cdot 10^4$ | 0.41 |
| | 10 | 5.43 | 1.29 | 6.16 | 5.07 | $2.04 \cdot 10^{-6}$ | $7.43 \cdot 10^7$ | 3156 |
| **Pro-tumor Immunity** | 1 | 3.21 | 0.33 | 1.34 | 8.52 | $1.43 \cdot 10^{-7}$ | $2.81 \cdot 10^7$ | 2669 |
| | 2 | 3.22 | 0.44 | 1.91 | 8.95 | $1.38 \cdot 10^{-7}$ | $3.24 \cdot 10^7$ | 985 |
| | 3 | 5.67 | 0.23 | 0.43 | 9.24 | $1.37 \cdot 10^{-7}$ | $4.50 \cdot 10^6$ | 1237 |
| | 4 | 3.48 | 0.55 | 2.55 | 9.43 | $1.28 \cdot 10^{-7}$ | $8.91 \cdot 10^7$ | 374 |
| | 5 | 4.49 | 0.21 | 0.77 | 35.74 | $4.17 \cdot 10^{-7}$ | $3.76 \cdot 10^5$ | 93 |
| | 6 | 3.10 | 0.34 | 1.39 | 8.70 | $1.46 \cdot 10^{-7}$ | $4.27 \cdot 10^7$ | 2914 |
| | 7 | 3.34 | 0.37 | 1.46 | 9.66 | $1.50 \cdot 10^{-7}$ | $2.55 \cdot 10^7$ | 667 |
| | 8 | 2.86 | 0.37 | 1.70 | 13.55 | $1.91 \cdot 10^{-7}$ | $2.46 \cdot 10^8$ | 466 |
| | 9 | 3.26 | 0.29 | 1.22 | 11.72 | $2.00 \cdot 10^{-7}$ | $3.17 \cdot 10^7$ | 2483 |
| | 10 | 3.04 | 0.33 | 1.39 | 11.28 | $1.71 \cdot 10^{-7}$ | $5.55 \cdot 10^7$ | 863 |

**Table 2.** Parameter values for immune predation of cancer cells, immune growth, and immune recruitment. The homeostatic immune presence parameter, $I_e$, is determined by the MCMC fitting method and is thus specific to each parameter set listed in Table 1.

| | | | |
|---|---|---|---|
| **Antitumor Immunity** | $a = \frac{2}{10}$ | $b = \frac{3}{10}$ | |
| **Pro-tumor Immunity** | $a = \frac{2}{10}$ | $b = \frac{1}{10}$ | |
| **Immune Predation** | $\theta = 2.5$ | $\phi = 50$ | $\beta = 0.5$ |
| **Immune Growth / Recruitment** | $\lambda = 0.2$ [day$^{-1}$] | $r = 0.001$ | |
| **Immune Carrying Capacity Regulation** | $x = 6$ [day$^{-1}$] | $y = 10^{-7}$ [(Cell No.)$^{-2/3}$ day$^{-1}$] | $z = 0.09$ [day$^{-1}$] |

## Appendix A1

Cancer-immune phase portraits demonstrate the variability across the 10 parameter sets in the antitumor immunity ensemble (values listed in Table 1). Tumors are simulated to grow from an initial injection of $C_0 = 10^4$ cancer cells and varying numbers of immune cells ($I_0 = \gamma C_0$). The ranges (values of $\gamma$) that divide the behavior between tumor growth (red) or tumor suppression (blue), are listed below the plots. Simulations result from solving the full system of equations (1)-(4) with direct predation through $\Psi$, equation (5). Axes indicate diameter of spherical population in mm.

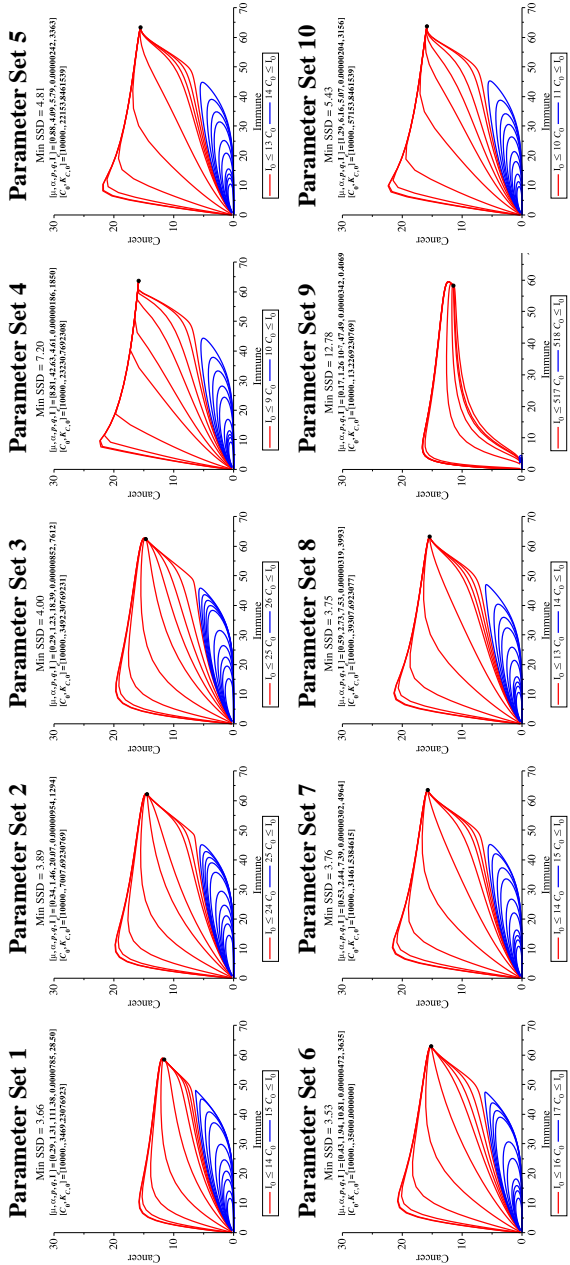

## Appendix A2

Cancer-immune phase portraits demonstrate the variability across the 10 parameter sets in the pro-tumor immunity ensemble (values listed in Table 1). Tumors are simulated to grow from an initial injection of $C_0 = 10^4$ cancer cells and varying numbers of immune cells ($I_0 = \gamma C_0$). The ranges (values of $\gamma$) that divide the behavior between tumor growth (red) or tumor suppression (blue), are listed below the plots. Simulations result from solving the full system of equations (1)-(4) with direct predation through $\Psi$, equation (5). Axes indicate diameter of spherical population in mm.

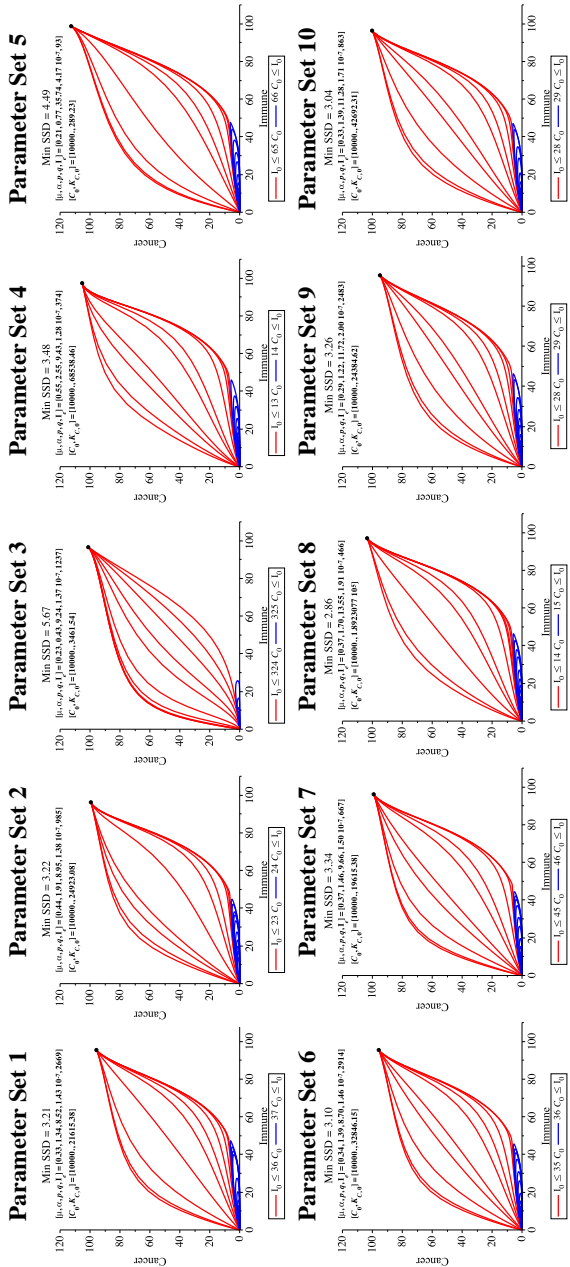